\documentclass[prd,aps,twocolumn,superscriptaddress,showpacs,amsmath,amssymb,floatfix,nofootinbib,10pt]{revtex4}

\usepackage{amsfonts}
\usepackage[dvipdfm,colorlinks=true, linkcolor=blue, citecolor=blue, urlcolor=blue]{hyperref}
\usepackage{color,xcolor,graphicx,amsfonts,amsmath,amssymb,multirow}
\usepackage{graphicx,amsfonts,multirow}
\usepackage{amsmath}

\newcommand{\DS}[1]{/\!\!\!#1}
\usepackage{epstopdf,multirow}
\usepackage{newtx}
\DeclareMathAlphabet{\mathcal}{OMS}{cmsy}{m}{n}
\usepackage{ulem}

\allowdisplaybreaks
\date{\today}
\begin{document}

\title{Probing $D_s^*$-meson longitudinal twist-2 LCDA}
\author{Si-Hai Zhang}
\address{Department of Physics, Guizhou Minzu University, Guiyang 550025, P.R.China}
\author{Tao Zhong}
\author{Hai-Bing Fu}
\email{fuhb@gzmu.edu.cn}
\address{Department of Physics, Guizhou Minzu University, Guiyang 550025, P.R.China}
\address{Institute of High Energy Physics, Chinese Academy of Sciences, Beijing 100049, P.R.China}
\author{Ya-Xiong Wang}
\author{Wan-Bing Luo}
\address{Department of Physics, Guizhou Minzu University, Guiyang 550025, P.R.China}
\begin{abstract}
In this paper, we carry on an investigation of the semileptonic decays $B_s\to D_s^*\ell \bar\nu_{\ell}$. Firstly, we derive the moments of the $D_s^*$-meson longitudinal leading-twist light-cone distribution amplitude (LCDA) based on QCD sum rules within background field theory framework. Considering the contributions of the vacuum condensates up to dimension-six, its first ten non-zero $\xi$-moments at the initial scale $\mu_0 =1~{\rm GeV}$ are $\langle \xi^{\|, 1}_{2; D_s^*} \rangle|_{\mu_0} = -0.328_{-0.051}^{+0.041}$, $\langle\xi^{\|, 2}_{2;D_s^*}\rangle|_{\mu_0} = +0.260_{-0.039}^{+0.045}$, $\langle\xi^{\|, 3}_{2;D_s^*}\rangle|_{\mu_0} = -0.130_{-0.020}^{+0.016}$, $\langle\xi^{\|, 4}_{2;D_s^*}\rangle|_{\mu_0} = +0.111_{-0.016}^{+0.018}$, $\langle\xi^{\|, 5}_{2; D_s^*} \rangle|_{\mu_0} = -0.071_{-0.011}^{+0.009}$, $\langle\xi^{\|, 6}_{2;D_s^*}\rangle|_{\mu_0} = +0.056_{-0.008}^{+0.009}$, $\langle\xi^{\|, 7}_{2;D_s^*}\rangle|_{\mu_0} = -0.044_{-0.007}^{+0.006}$, $\langle\xi^{\|, 8}_{2;D_s^*}\rangle|_{\mu_0} = +0.039_{-0.006}^{+0.007}$, $\langle\xi^{\|, 9}_{2; D_s^*} \rangle|_{\mu_0} = -0.027_{-0.004}^{+0.003}$ and $\langle\xi^{\|, 10}_{2;D_s^*}\rangle|_{\mu_0} = +0.028_{-0.004}^{+0.005}$, respectively. Meanwhile, we construct the $D_s^*$-meson longitudinal leading-twist LCDA by using the light-cone harmonic oscillator model. Then, using those moments, we fix the model parameters $\alpha_{2;D_s^*}$ and $B_1^{2;D_s^*}$ by the least square method and apply them to calculate $B_s \to D_s^*$ transition form factors $A_1(q^2), A_2(q^2)$ and $V(q^2)$ that are derived by using the QCD light-cone sum rules. At the large recoil region, we obtain $A_1(0) =0.632_{-0.135}^{+0.228}, A_2(0) =0.706_{-0.092}^{+0.109}$ and $V(0) =0.647_{-0.069}^{+0.076}$. Those form factors are then extrapolated to the allowed whole physical $q^2$-region through the simplified series expansion. Finally, we obtain the branching fractions for the two decay channels $B_s\to D_s^*\ell\bar\nu_\ell$, $\it i.e.$ ${\cal B}(B_s^0 \to D_s^{*+}e^-\bar\nu_e)=(5.45_{-1.57}^{+2.15})\times 10^{-2}$, ${\cal B}(B_s^0 \to D_s^{*+}\mu^-\bar\nu_\mu)=(5.43_{-1.57}^{+2.14})\times 10^{-2}$.
\end{abstract}
\maketitle

\section{Introduction}
$B$-meson semileptonic decay is one of the very important tools for studying the weak decay interaction. It have great phenomenological implications within the Standard Model (SM) of particle physics. 
The $B_s\to D_s^*\ell \bar\nu_{\ell}$ decay provides an opportunity for extracting CKM matrix elements and testing the SM~\cite{BaBar:2012obs, Belle:2015qfa, Belle:2019rba, HFLAV:2019otj}. Recently, the study of $B_s\to D_s^*\ell \bar\nu_{\ell}$ has attracted significant interest, driven by advancements in experimental capabilities and theoretical developments. Many experiments have provided more and more precise data of these decays, which allow the extraction of the CKM matrix element to an increasingly better accuracy. 
In addition, many theoretical groups have also generated great interest in exploring the decay channel.

In 2020, the LHCb Collaboration~\cite{LHCb:2020cyw,LHCb:2020hpv} published an article on the experimental measurement of $B_s\to D_s^*\ell \bar\nu_{\ell}$ decay. They used the experimental data samples collected by the LHCb detector at the center-of-mass energies of 7 and 8 TeV and conducted systematic studies on the two decay channels $B^0_s\to D^-_s\mu^+\nu_\mu $ and $B^0_s\to D^{*-}_s\mu^+\nu_\mu $ using Caprini-Lellouch-Neubert (CLN) and Boyd-Grinstein-Lebed (BGL) parameterizations. Finally, the measured values of $|V_{cb}|$ obtained under the two parameterizations are $(41.4 \pm 0.6 \pm 0.9 \pm 1.2)\times10^{-3}$  and $(42.3 \pm 0.8\pm 0.9 \pm 1.2)\times 10^{-3}$, respectively. In reference~\cite{Harrison:2023dzh}, the lattice QCD and HPQCD Collaboration worked together to determine the model independent value of $V_{cb}= 39.03(56)_{\rm exp.}(67)_{\rm latt.}\times 10^{-3}$ using the $B\to D^{*}\ell \bar{\nu}_{\ell}$ data from Belle and the $B_s\to D_s^*\mu \bar{\nu}_\mu $ data from LHCb, as well as their own transition form factors (TFFs). In addition, the TFFs and decay width  behaviors for the $B_s\to D_s^*\ell \bar{\nu}_{\ell}$ process are also given in the article.

Theoretically, there are many theoretical methods for making reasonable theoretical predictions of $B_s\to D_s^*\ell \bar\nu_{\ell}$ decay, such as lattice QCD (LQCD)~\cite{Harrison:2021tol, Penalva:2023snz, Harrison:2017fmw, McLean:2019sds, Bordone:2019guc}, covariant confined quark model (CCQM)~\cite{Soni:2021fky}, QCD sum rules (QCDSR)~\cite{Blasi:1993fi}, relativistic quark model (RQM)~\cite{Faustov:2012mt}, perturbative QCD (pQCD)~\cite{Fan:2013kqa, Hu:2019bdf}, light-front quark model (LFQM)~\cite{Li:2010bb}, and bethe-salpeter method~\cite{Zhou:2019stx}. Then we provides a brief introduction to the above methods. In Ref.~\cite{Harrison:2021tol}, the SM semileptonic vector and axial-vector form factors for $B_s \to D_s^*$ decay are calculated via the LQCD. The relevant calculation methods, the dependence on the heavy quark mass, decay rates and ratios are analysed. Meanwhile, the consistency with LHCb results and the tests on the impact of new physics couplings are also discussed. They give a reasonable reference to study the semileptonic decay of $B_s \to D_s^* \ell \bar\nu_{\ell}$.
In Ref.~\cite{Soni:2021fky}, the author relies on the Standard Model framework based on the CCQM to carry out the calculation of TFFs within the entire dynamical category of squared momentum transfer for semileptonic decays. The obtained results, such as decay width ratios, show a degree of consistency with LHCb experiments and LQCD simulations, and the behaviors of differential decay distributions are compared. Meanwhile, other physical observations are also calculated. Through calculation, $R(D_s) = 0.271\pm0.069$, $R(D^*_s) =0.240\pm0.038$ are obtained, and the ratio of decay widths of the $D_s$ and $D_s^*$ channels in the muon meson mode $\Gamma(B_s\to D_s\mu^+\nu_\mu)/\Gamma(B_s\to D_s^*\mu^+\nu_\mu)=0.451\pm0.093$ is determined. In Ref.~\cite{Blasi:1993fi}, the phenomenology of the $b$-flavored strange meson $B^0_s$ is studied through the QCDSR. Specifically, this includes the evaluation of the particle's mass and leptonic constant, as well as the study of the form factors of certain decays (such as $\bar B^0_s \to  D^+_s\ell^-\bar{\nu}$, $\bar B^0_s\to  D^{*+}_s\ell^-\bar{\nu}$, $\bar B^0_s \to K^{*+}\ell^-\bar{\nu}$); at the same time, the calculation of two-body non-leptonic $\bar B^0_s$ decays is carried out under the factorization approximation; finally, the evaluation result of the $SU(3)_F$ breaking effect in the $\bar B^0_s$ channel is compared with other estimates.

Also as a well-established theory that can be effectively applied to the exclusive decay process, the QCD light-cone sum rule (LCSR)~\cite{Ball:1991bs,Chernyak:1990ag} incorporates both the hard and soft contribution in the computation of hadron transitions. In the LCSR method, the two-point correlation function of vacuum to meson is constructed for calculating heavy-to-light TFFs. In which, the matrix elements of nonlocal operators is carried out at light-cone region $x^2 \to 0$. The difference between this method and the traditional SVZ sum rule~\cite{Shifman:1978by} is that the all non-perturbative dynamics are parameterized according to the light-cone distribution amplitudes (LCDAs) with progressively higher twists instead of quark and gluon condensates ~\cite{Khodjamirian:2011jp}. Currently, the LCSR has been widely applied to the study of $B$ light-flavor meson decays~\cite{Ball:2005tb, Yang:2024jlz, Duplancic:2008ix, Khodjamirian:2011ub, Wu:2009kq,Yang:2024ang}. In this work, we will adopt LCSR to calculate the $B_s\to D^*_s$ TFFs. Specifically, due to the presence of both longitudinal $(\|)$ and transverse $(\perp)$ polarization states in vector mesons, employing traditional currents to construct correlation functions poses a challenge involving fifteen LCDAs. Therefore, to simplify the calculation process, we will adopt chiral currents instead of traditional currents, allowing the contributions of twist-2 LCDA to dominate and eliminating contributions from other LCDAs. The specific procedure will be elaborated in the next section. Therefore, an accurate prediction of the twist-2 $\phi_{2;D_s^*}^\|(x,\mu)$ is of great importance. So far, the LCDA of many mesons depends on Gegenbauer moment, which can be calculated using additional sum rules. As a mature theoretical method, the background field theory (BFT) decomposes the quark field into a classical background field that describes non-perturbative effects and a quantum field that describes perturbative effects. This can provide a clean physical picture for separating the perturbative and non-perturbative properties of the QCD theory and provide a systematic way to
derive the QCDSR for hadron phenomenology. Meanwhile, due to the ability to adopt different gauges for quantum fluctuations and the background field in BFT, the calculations can be greatly simplified. In our previous work, the longitudinal twist-2 LCDA of $\rho$-meson is successfully studied in BFT and constructed by light cone harmonic oscillator (LCHO) model~\cite{Fu:2016yzx}.  Motivated by this, in this work, we will employ the BFT method to study the twist-2 LCDA of $D_s^*$-meson and attempt to integrate the phenomenological LCHO model to provide us with another perspective to understand the momentum distribution of quarks and gluons inside $D_s^*$-mesons.

The rest of this article are organized as follows. In Sec.~\ref{Sec:II}, we derive the summation rules for the $\xi$-moments of the $D_s^*$-meson longitudinal leading-twist $\phi_{2;D_s^*}^\|(x,\mu)$ LCDAs and the $B_s\to D_s^*$ TFFs $A_1(q^2)$, $A_2(q^2)$ and $V(q^2)$, and established the LCHO for the $D_s^*$-meson leading-twist LCDAs. In Sec.~\ref{Sec:III}, we provide relevant numerical results and make a detailed discussion. Section~\ref{Sec:IV} is reserved for summary.

\section{Theoretical Framework}\label{Sec:II}
The differential decay width of semileptonic decays $B_s\to D_s^*\ell\bar\nu_\ell$ can be written in terms of the helicity components basis \cite{Zhang:2020dla, Ivanov:2015tru, Ivanov:2019nqd}:
\begin{eqnarray}
\frac{d\Gamma(B_s\to D_s^*\ell\bar\nu_\ell)}{dq^2} =\frac{G_F^2|V_{cb}|^2 \lambda^{1/2} q^2}{192 \pi^3 m_{B_s}^3}\left(1-\frac{m_\ell^2}{q^2}\right){\cal H}_{\rm total},
\label{Eq:DecayWidth}
\end{eqnarray}
where the Fermi coupling constant $G_F =1.1663787(6)\times 10^{-5}~{\rm GeV}^{-2}$, $|V_{cb}|$ is the CKM matrix element, $\lambda\equiv\lambda(m_{B_s}^2,m_{m_{D_s^*}}^2,q^2) =m_{B_s}^4+m_{D_s^*}^4+q^4-2(m_{B_s}^{2}m_{D_s^*}^{2}+m_{B_s}^{2}q^{2}+m_{D_s^*}^{2}q^{2})$ is the phase-space factor, $m_\ell$ stands for the lepton mass $(\ell =e,\mu,\tau)$, and ${\cal H}_{\rm total}$ represents the overall helicity structure:
\begin{eqnarray}
{\cal H}_{\rm total} =({\cal H}_U +{\cal H}_L)\left(1 +\frac{m_\ell^2}{2q^2}\right) +\frac{3m_\ell^2}{2q^2}{\cal H}_S.
\label{Eq:Htotal}
\end{eqnarray}
The symbols ${\cal H}_{I}(I =U,L,S)$ are the bilinear combinations of the helicity components of the hadronic tensor. Here, in this work the leptonic mass $m_\ell$ is very small in case of $\ell=(e,\mu)$ when compared with the squared transition momentum $q^2$, which can be safely neglected. Thus there will only leaves two helicity structures, {\it i.e.} ${\cal H}_U$ and ${\cal H}_L$ which have the following formulas
\begin{eqnarray}
{\cal H}_U =|H_+|^2 +|H_-|^2, \quad {\cal H}_L =|H_0|^2.
\label{Eq:HULS}
\end{eqnarray}
The helicity amplitudes $H_i$ with the index $i=(\pm,0,t)$ are denoted as the function of invariant mass $q^2$, which are formed from the $B_s\to D_s^*$ TFFs with different combinations:
\begin{align}
H_{\pm}(q^2)&=(m_{B_s}+m_{D_s^*})A_1(q^2)\mp\frac{\lambda^{1/2}}{m_{B_s}+m_{D_s^*}}V(q^2),
\nonumber\\
H_{0}(q^2)&=\frac{1}{2m_{D_s^*}\sqrt{q^2}}\bigg[ (m_{B_s}+m_{D_s^*})(m_{B_s}^2-m_{D_s^*}^2-q^2)
\nonumber\\
&\times A_1(q^2)-\frac{\lambda}{m_{B_s}+m_{D_s^*}}A_2(q^2) \bigg],
\label{Eq:HUL}
\end{align}

As we know, the three $B_s\to D_s^*$ TFFs $A_1(q^2)$, $A_2(q^2)$ and $V(q^2)$ are the important hadronic inputs for studying the relevant implications of semileptonic decays $B_s\to D_s^*\ell\bar\nu_\ell$. Therefore, to derive their analytic expressions within the LCSR approach, we construct the following chiral current correlation function (correlator):
\begin{eqnarray}
\Pi_\mu(p,q) =i\int d^4xe^{iq\cdot x}\langle D_s^*(p)|T\{J_\mu(x),J_{B_s}^\dagger(0)\}|0\rangle, \label{correlator}
\end{eqnarray}
where $J_\mu(x) = \bar c(x)\gamma_\mu(1+\gamma_5)b(x)$ and $J_{B_s}^\dagger(0) =\bar b(0)i(1+\gamma_5)s(0)$ is the left-handed current. As the $D_s^*$-meson DAs are relatively complex structures, there are both chiral-even and chiral-odd DAs for $D_s^*$-meson, which has longitudinal and transverse polarization states. The adopted left-handed current can effectively highlight the contributions from the chiral-even DAs such as  $\phi_{2;D_s^*}^\|(x,\mu), \phi_{3;D_s^*}^{\bot}(x,\mu), \psi_{3;D_s^*}^{\bot}(x,\mu), \Phi_{3;D_s^*}^\|(x,\mu), \tilde{\Phi}_{3;D_s^*}^\|, \phi_{4;D_s^*}^\|(x,\mu)$, and $\psi_{4;D_s^*}^\|(x,\mu)$, while the chiral-odd DAs provide zero contributions. For the remaining chiral-even DAs, only the  $\phi_{2;D_s^*}^\|, \phi_{3;D_s^*}^\bot(x,\mu)$, and $\psi_{3;D_s^*}^\bot(x,\mu)$ account for dominant contributions to the LCSR, while other chiral-even DAs offer negligible contributions. Moreover, the twist-3 DAs $\phi_{3;D_s^*}^\bot(x,\mu)$ and $\psi_{3;D_s^*}^\bot(x,\mu)$ can be linked with $\phi_{2;D_s^*}^\|(x,\mu)$ under the Wandzura-Wilczek (WW) approximation ~\cite{Ball:1997rj,Wandzura:1977qf}:
\begin{align}
&\phi_{3;D_s^*}^{\bot;{\rm WW}}(x,\mu) =\frac{1}{2}\left[ \int_0^x \!\! dv\frac{\phi_{2;D_s^*}^\|(v,\mu)}{v} \! + \! \int_x^1dv\frac{\phi_{2;D_s^*}^\|(v,\mu)}{v}\right],
\nonumber\\
&\psi_{3;D_s^*}^{\bot;{\rm WW}}(x,\mu) =2\!\left[
\bar x \! \int_0^x \!\!\! dv  \frac{\phi_{2;D_s^*}^\|(v,\mu)}{v} \!+\!x\!\int_x^1\!\!dv\frac{\phi_{2;D_s^*}^\|(v,\mu)}{v}\!
\right].
\end{align}
Therefore, the longitudinal leading-twist $\phi_{2;D_s^*}^\|(x,\mu)$ may provide a dominant contribution, either directly or indirectly.

The correlation function~\eqref{correlator} is defined at both the time-like and the space-like $q^2$-regions. In the time-like $q^2$-region, long distance quark-gluon interactions are dominant. The correlator can insert a complete set of the $B_s$-meson states with the same $J^P$ quantum numbers to acquire the hadron expression in physical region. Meanwhile, in space-like region, the correlator can be treated by the OPE in deep Euclidean region with the coefficients being pQCD calculable. After matching the two results by the dispersion relation, the final LCSR can be obtained by applying the Borel transformation, which is used to suppress the less known continuum contributions, and subsequently the resultant $B_s\to D_s^*$ TFFs under the LCSR approach read off:
\begin{align}
A_1(q^2)&=\frac{2m_{b}^{2}m_{D_s^*}f_{D_s^*}^\|} {f_{B_s}m_{B_s}^{2}(m_{B_s} \! +\! m_{D_s^*})e^{-m_{B_s}^{2}/M^2}}\Bigg\{\!\!  \int_0^1\! {\frac{du}{u}}e^{-s(u)/M^2}
\nonumber\\
& \times\left[ \Theta (c(u,s_0))\phi_{3;D_s^*}^{\bot}(u)-\frac{m_{D_s^*}^{2}}{u M^2}\tilde{\Theta}(c(u,s_0)) C_{D_s^*}^\|(u) \right]
\nonumber \\
&-m_{D_s^*}^{2}\,\int{D}\,\underline{\alpha}\,\int dv \,e^{-s(X)/M^2} \, \frac{1}{X^2 M^2} \Theta(c(X,s_0))
\nonumber\\
& \times \left[\Phi_{3;D_s^*}^\|(\underline{\alpha})+\tilde{\Phi}_{3;D_s^*}^\|(\underline{\alpha})\right] \Bigg\},
\nonumber\\
\label{Eq:TFFA1q2}
\\
A_2(q^2)&=\frac{m_{b}^{2}m_{D_s^*}(m_{B_s} \! + \! m_{D_s^*}) \! f_{D_s^*}^\|} {f_{B_s}m_{B_s}^{2}e^{-m_{B_s}^{2}/M^2}} \!\Bigg\{ 2 \!\! \int_0^1 \! {\frac{du}{u}}e^{-s(u)/M^2} \!\bigg[ \frac{1}{uM^2}
\nonumber\\
& \times \, \tilde{\Theta} \, (c(u,s_0))\,A_{D_s^*}^\|(u) \,+\, \frac{m_{D_s^*}^{2}}{u \, M^4} \, \tilde{\tilde{\Theta}} \,(c(u,s_0)) C_{D_s^*}^\|(u)
\nonumber\\
&+\frac{m_{b}^{2}m_{D_s^*}^{2}}{4u^4M^6}\tilde{\tilde{\tilde{\Theta}}}(c(u,s_0))B_{D_s^*}^\|(u) \bigg] + m_{D_s^*}^{2} \int\! \mathcal{D} \underline\alpha  \int dv
\nonumber\\
&\times \frac{e^{-s(X)/M^2}}{X^3 M^4} \Theta (c(X,s_0)) [\Phi_{3;D_s^*}(\underline{\tilde{\alpha}}) + \tilde{\Phi}_{3;D_s^*}^\|(\underline{\tilde{\alpha}})] \Bigg\},
\label{Eq:TFFA2q2}
\\
V(q^2)&=\frac{m_{b}^{2}m_{D_s^*}(m_{B_s}+m_{D_s^*})f_{D_s^*}^\|}
{2f_{B_s}m_{B_s}^{2}e^{-m_{B_s}^{2}/M^2}}\int_0^1{du}e^{-s(u)/M^2}\frac{1}{u^2 M^2}
\nonumber\\
&\times\tilde{\Theta}(c(u,s_0))\psi_{3;D_s^*}^{\bot}(u).
\label{TFF:V}
\end{align}
The three simplified $D_s^*$-meson LCDAs $A_{D_s^*}^\|(u), B_{D_s^*}^\|(u)$, and $C_{D_s^*}^\|(u)$ are represented as follows, respectively:
\begin{align}
    &A_{D_s^*}^\|(u) =\int_0^udv\left[\phi_{2;D_s^*}^\|(v) -\phi_{3;D_s^*}^{\bot}(v) \right],
    \\
    &B_{D_s^*}^\|(u) =\int_0^udv\phi_{4;D_s^*}^\|(v),
    \\
    &C_{D_s^*}^\|(u) =\int_0^udv\int_0^vdw\bigg[\psi_{4;D_s^*}^\|(w) +\phi_{2;D_s^*}^\|(w)
    \nonumber
    \\
    &\hspace{1.1cm}-2\phi_{3;D_s^*}^{\bot}(v)\bigg].
\end{align}
Furthermore, $s(\varrho) =[m_b^2 -\bar \varrho(q^2-\varrho m_{D_s^2}^2)]/\varrho$ with $\bar \varrho = (1-\varrho)$ and $\varrho$ refers to $u$ or $X = (a_1 +va_3)$. $f_{D_s^*}^\|$ is the $D_s^*$-meson decay constant. The usual step function $\Theta(c(\varrho,s_0))$, where $c(\varrho,s_0) =us_0-m_b^2 +\bar uq^2 -u\bar um_{D_s^*}^2$ is defined such that if $c(\varrho,s_0) <0$, it is zero; otherwise, it is 1. The definitions of $\tilde{\Theta}(c(u,s_0))$ and $\tilde{\tilde{\Theta}}(c(u,s_0))$ are
\begin{align}
    &\int_0^1\frac{du}{u^2M^2}e^{-s(u)/M^2}\tilde{\Theta}(c(u,s_0))f(u)
    \nonumber\\
    & \qquad =\int_{u_0}^1\frac{du}{u^2M^2}
    e^{-s(u)/M^2}f(u)+\delta(c(u_0,s_0)),
    \\
    &\int_0^1\frac{du}{2u^3M^4}e^{-s(u)/M^2}\tilde{\tilde{\Theta}}(c(u,s_0))f(u)
    \nonumber\\
    &\qquad =\int_{u_0}^1\frac{du}{2u^3M^4}e^{-s(u)/M^2}f(u)+\Delta(c(u_0,s_0)),
\end{align}
where $\delta(c(u,s_0))=e^{-s_0/M^2}f(u_0)/C_0$ with $C_0=m_b^2+u_0^2m_{D_s^*}^2-q^2$and
\begin{align}
\Delta(c(u,s_0))=e^{-s_0/M^2}\!\left[ \frac{1}{2u_0M^2}\frac{f(u_0)}{C_0}
-\frac{u_0^2}{2C_0}\frac{d}{du}\bigg(\frac{f(u)}{uC}\bigg)\Big|_{u=u_0} \right].
\end{align}

At the stage, we shall focus on the longitudinal leading-twist DAs of $D_s^*$-meson. According to our previous works~\cite{Fu:2018vap,Hu:2024tmc,Zhong:2023cyc}, which show the technique for constructing the LCHO model of vector mesons's wavefuntion (WF) based on the Brodsky-Huang-Lepage (BHL)~\cite{Brodsky:1983st,Brodsky:1983vf}, one can obtain
\begin{align}
    \Psi_{2;D_s^*}^\|(x,{\bf k}_\bot) &=\sum_{\lambda_1\lambda_2}\chi_{2;D_s^*}^{\lambda_1\lambda_2}(x,{\bf k}_\bot)\Psi_{2;D_s^*}^R(x,{\bf k}_\bot)
    \nonumber\\
    &=\frac{\tilde{m}}{\sqrt{{\bf k}_\bot^2 +\tilde{m}^2}}A_{2;D_s^*}\varphi_{2;D_s^*}(x)
    \nonumber\\
    &\times \exp\left[
        -\frac{1}{\beta_{2;D_s^*}^2}\left(
            \frac{{\bf k}_\bot^2 +\hat m_c^2}{1-x} +\frac{{\bf k}_\bot^2 +\hat m_s^2}{x}
        \right)
    \right],
\end{align}
where equivalent mass $\tilde{m} =\hat m_cx +\hat m_s(1-x)$ with $\hat m_{c}$ and $\hat m_{s}$ refer to the constituent quark masses and $\lambda_{1(2)}$ are the helicities of two constituent quarks. $A_{2;D_s^*}$ is the normalization constant, $\beta_{2;D_s^*}$ govens the transverse distribution, and $\varphi_{2;D_s^*}(x) =1 +B_1^{2;D_s^*}C_1^{3/2}(2x-1)$ governs the longitudinal distribution. By considering the relation between the leading-twist DA and its WF,
\begin{eqnarray}
    \phi_{2;D_s^*}^\| (x,\mu_0^2) = \frac{{2\sqrt 6 }}{{f_{D_s^*}^\|}} \int_{|{\bf{k}}_\bot|^2 \le \mu_0 ^2 } {\frac{{d^2{\bf{k}}_ \bot  }}{{16\pi ^3 }} \Psi_{2;D_s^*}^\| (x,{\bf{k}}_\bot )},
    \label{eq:LWF}
\end{eqnarray}
one can derive the leading-twist DA $\phi_{2;D_s^*}^\| (x,\mu_0^2)$ aftering integrating the transverse momentum ${\bf k}_\bot$:
\begin{align}
    \phi_{2;D_s^*}^\| (x,\mu_0^2) & = \frac{\sqrt{6}A_{D_s^*}\beta_{2;D_s^*}^2}{\pi ^{2}f_{D_s^*}^\|}x(1-x)\varphi_{2;D_s^*}^\|(x,\mu)
    \nonumber\\
    &\times\exp \biggl[-\frac{x\hat{m}_{c}^{2}+(1-x)\hat{m}_{s}^{2}}{8\beta_{2;D_s^*}^{2}x(1-x)} \biggr]
    \nonumber\\
    &\times\Bigg\{1-\exp \biggl[ -\frac{\mu_0^2}{8\beta_{2;D_s^*}^{2}x(1-x)} \biggl]\Bigg\},
    \label{eq:DA}
\end{align}
where $\mu_0\sim \Lambda_{\rm QCD}$ represents the factorization scale. Additionally, the model parameters $A_{2;D_s^*},\beta_{2;D_s^*}$, and $B_1^{2;D_s^*}$ can be appropriately determined by relative constraints, such as the $\phi_{2;D_s^*}^\|(x,\mu_0^2)$ normalization condition, the Fock-state expansion of the $D_s^*$-meson, and the Gegenbauer moments of $\phi_{2;D_s^*}^\|(x,\mu_0^2)$~\cite{Fu:2018vap,Hu:2024tmc,Zhong:2023cyc}.

On the other hand, considering the definitions of the leading-twist DA $\phi_{2;D_s^*}^\| (x,\mu)$,
\begin{align}
    &\langle 0|\bar c(z)\DS{z}\gamma_5s(-z)|D_s^*(q)\rangle =i(z\cdot q)f_{D_s^*}^\|
    \nonumber\\
    &\qquad \times \int_0^1dx e^{i(2x-1)(z\cdot q)}\phi_{2;D_s^*}^\| (x,\mu), \label{DA_definition}
\end{align}
and
\begin{align}
    &\langle \xi_{2;D_s^*}^{\|;n}\rangle|_\mu =\int_0^1dx(2x-1)^n \phi_{2;D_s^*}^\| (x,\mu),
    \\
    &\langle \xi_{2;D_s^*}^{\|;0}\rangle|_\mu =\int_0^1dx\phi_{2;D_s^*}^\| (x,\mu) =1,
\end{align}
one can consider the following two-point correction function to derive the $n$th-order $\xi$-moments of $D_s^*$-meson,
\begin{align} \nonumber
    \Pi_{2;D_s^*}^{(n,0)}(z,q) &=i\int d^4x^{iq\cdot x}\langle 0|T\{J_n(x)J_0^\dagger(0)\}|0\rangle
    \\
    &=(z\cdot q)^{n+2}I_{2;D_s^*}^{(n,0)}(q^2).\label{DA_correlator}
\end{align}
\begin{figure}[b]
\begin{center}
\includegraphics[width=0.38\textwidth]{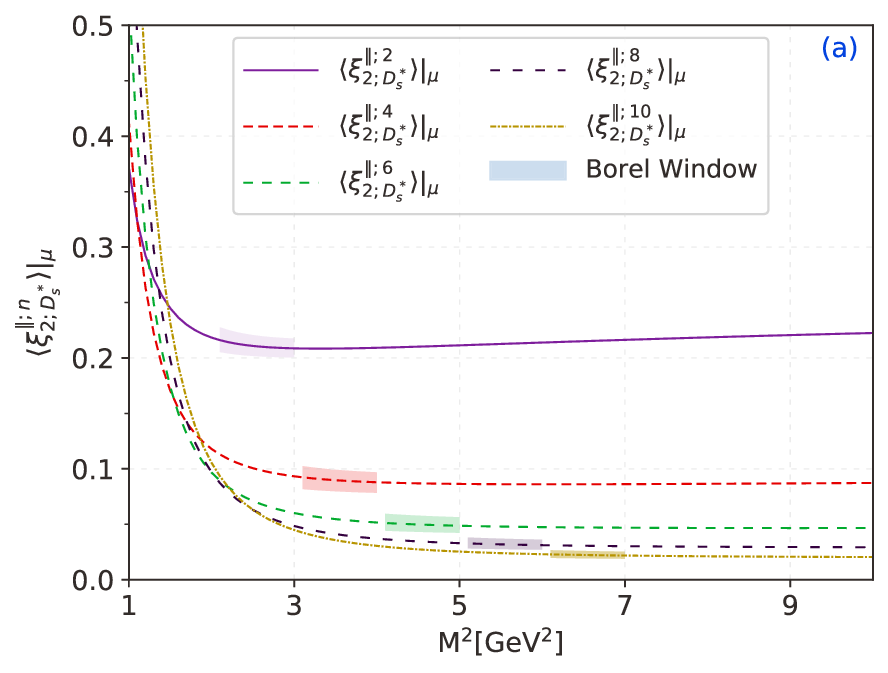}
\includegraphics[width=0.38\textwidth]{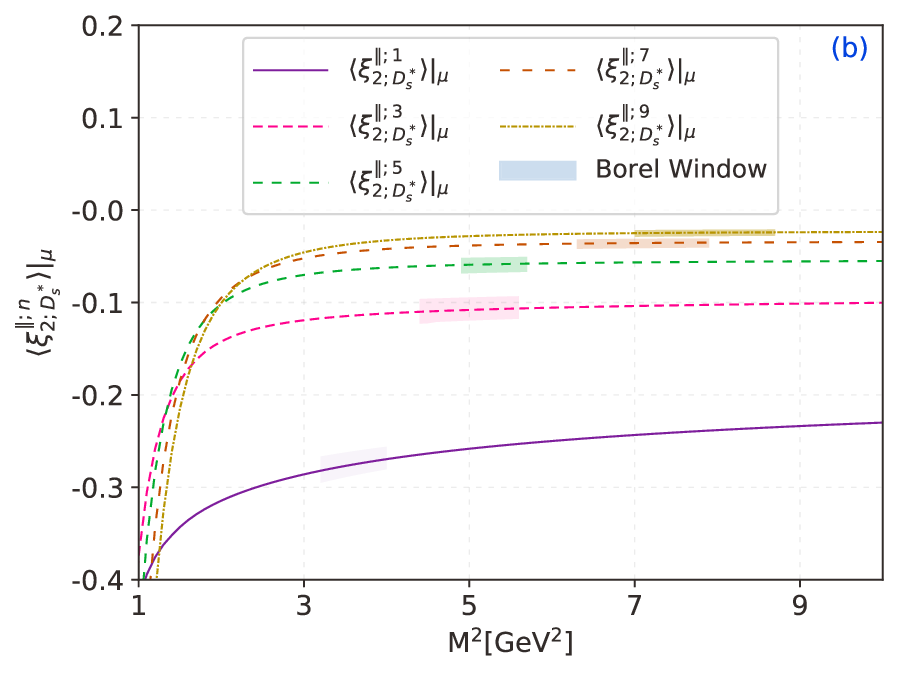}
\end{center}
\caption{The $D_s^*$-meson twist-2 LCDA moments $\langle\xi^{\|;n}_{2;D_s^*}\rangle|_\mu$ with $(n=1,...,10)$ versus the Borel parameter $M^2$. Here, we only present the central values to show the curves of different moments clearly.}
\label{Fig:fp1}
\end{figure}
The currents $J_n(x) =\bar c(x)\DS{z}\gamma_5(iz\cdot\overleftrightarrow{D})^ns(x)$ and $J_0^\dagger(0) =\bar s(0)\DS{z}\gamma_5c(0)$ can be derived from Eq.~(\ref{DA_definition}) by expanding the its l.h.s. around $z \rightsquigarrow 0$ and expanding the exponential on the r.h.s. of Eq.~(\ref{DA_definition}) into a power series. Subsequently, according to the method of BFTSR~\cite{Huang:1989gv}, one can get the analytical expression of the $D_s^*$-meson longitudinal leading-twist DA moments $\langle \xi_{2;D_s^*}^{\|,n}\rangle|_\mu$:
\begin{align}
    \langle\xi^{\|;n}_{2;D_s^*}\rangle|_\mu \!=\!\frac{M^2 \! e^{m_{D_s^*}/M^2}}{f_{D_s^*}^{\|2}}&\Bigg\{\,\frac{1}{\pi \, M^2} \,
    \int_{t_{\rm min}}^{s_0^{D_s^*}} \, ds \, e^{-s/M^2} \,{\rm Im} \,I_{\rm pert}(s)
    \nonumber\\
    & +\hat L_MI_{\langle\bar qq\rangle}(-q^2) \, + \, \hat L_M \, I_{\langle G^2\rangle} \hspace{0.02cm} (-q^2)
    \nonumber\\
    & +\hat L_MI_{\langle\bar qGq\rangle}(-q^2) \!
    + \! \hat L_MI_{\langle \bar qq\rangle^2}(-q^2)
    \nonumber\\
    & +\hat L_MI_{\langle G^3\rangle}(-q^2) \Bigg\},
    \label{eq:Xi}
\end{align}
The perturbative and each condensates terms with Borel transformation have the following forms
\begin{widetext}
\begin{align}
    &{\rm Im}I_{\mathrm{pert}}(s)=\frac{3}{8\pi(n+1)(n+3)}\Bigg\{\bigg[2(n+1)\frac{m_{c}^{2}}{s}
    \left(1-\frac{m_{c}^{2}}{s}\right)+1\bigg]\left(1-\frac{2m_{c}^{2}}{s}\right)^{n+1}+(-1)^n\Bigg\},
    \label{Impert} 
    \\
    &\hat{L}_M I_{\langle \bar{s}s\rangle}(-q^2) = (-1)^n \exp \left( - \frac{m_c^2}{M^2} \right) \frac{m_s \langle  \bar ss\rangle}{M^4}, 
    \label{qq} 
    \\
    &\hat{L}_M I_{\langle G^2\rangle}(-q^2) = \frac{\langle \alpha_sG^2\rangle}{M^4} \frac{1}{12\pi} \bigg[ 2n(n-1) \mathcal{H}(n-2,1,1)
    + \mathcal{H}(n,0,0) - \frac{m_c^2}{M^2} \mathcal{H}(n,1,-2) \bigg], 
    \label{GG} 
    \\
    &\hat{L}_M I_{\langle \bar{s}Gs\rangle}(-q^2) = (-1)^n \exp \left(-\frac{m_c^2}{M^2}\right)\frac{m_s\langle g_s\bar{s}\sigma TGs \rangle}{M^6} 
    \left( -\frac{8n+1}{18} - \frac{2m_c^2}{9M^2} \right), 
    \label{qGq}
    \\
    &\hat{L}_M I_{\langle \bar{s}s\rangle^2}(-q^2) = (-1)^n \exp \left(-\frac{m_c^2}{M^2}\right) \frac{\langle g_s\bar{s}{s}\rangle^2}{M^6} \frac{2(2n+1)}{81}, 
    \label{qq2} 
    \\
    &\hat{L}_M I_{\langle G^3\rangle}(-q^2) = \frac{\langle g_s^3fG^3\rangle}{\pi^2 M^6} \, \bigg\{ \,\exp\left( -\frac{m_c^2}{M^2} \right) \, \bigg\{ -\frac{17}{96} \mathcal{F}_1(n,5,3,2,\infty) \, + \,\frac{n}{144} \mathcal{F}_2(n-1,5,3,1,\infty) \,- \,\frac{1}{96} \mathcal{F}_2(n,4,3,1,\infty) 
    \nonumber\\
    &\qquad   + \frac{1}{144}\mathcal{F}_2(n,3,3,1,\infty) - \frac{17}{96} \mathcal{G}_1(n,5) - \frac{17}{32} \mathcal{G}_2(n,5) \left( 1 - \frac{1}{3} \frac{m_c^2}{M^2} \right) + \frac{n}{144} \mathcal{G}_2(n-1,5) - \frac{n}{96} \mathcal{G}_3(n,4) + \frac{n}{144} \mathcal{G}_3(n,3)
    \nonumber\\
    &\qquad  + \frac{1}{288} \left[ 204\delta^{n0} + 204\theta(n-1)(-1)^n + (-1)^n \left( 100n - 154 + 51\frac{m_c^2}{M^2} \right) \right] \left[ \ln \frac{M^2}{\mu^2} + \psi(3) \right] + \frac{(-1)^n}{288} \left( 17\frac{m_c^2}{M^2} - 4n \right) \bigg\}
    \nonumber\\
    &\qquad  + \bigg\{\frac{1}{288} \Big[\! -4(n \! + \! 1)n(n \! - \!1) \mathcal{H}(n\!-\!2,1,1) \!+\! 4(n\!+\!1) \mathcal{H}(n,0,0) \!-\!  2n \mathcal{H}(n-1,1,-1) \!-\! 3 \mathcal{H}(n,0,-1) -\! 51 \mathcal{H}(n,1,-2) \bigg]
    \nonumber\\
    &\qquad  + \frac{1}{288} \frac{m_c^2}{M^2} \left[ -4n(n - 1) \mathcal{H}(n-2,1,0) 
    2 \mathcal{H}(n,0,-2) +\! 4 \mathcal{H}(n,0,-1) - \!2 \mathcal{H}(n-1,1,-2) - \!3 \mathcal{H}(n,1,-3) \right] + \!\frac{1}{240} \frac{m_c^4}{M^4} 
    \nonumber\\
    &\qquad \times \mathcal{H}(n,1,-4) \bigg\}\bigg\}. \label{GGG}
\end{align}
\end{widetext}
In which, the definition about the hypergeometric function $\mathcal{F}_{1,2}(n,a,b,l_{\rm min},l_{\rm max})$, $\mathcal{G}_{1,2,3}(n,a)$, $\mathcal{H}(n,a,b)$ and the traditional Borel transformation formulas $\hat{L}_M \frac{1}{(-q^2 + m_c^2)^k} \ln \frac{-q^2 + m_c^2}{\mu^2}$, $\hat{L}_M (-q^2 + m_c^2)^k \ln \frac{-q^2 + m_c^2}{\mu^2}$, $\hat{L}_M \frac{(-q^2)^l}{(-q^2+m_c^2)^{l+\tau}}$ can be found in our previous work~\cite{Zhong:2018exo, Zhang:2021wnv, Zhang:2017rwz}. Here, we have a notation that these functions have similar expressions in our previous work~\cite{Zhang:2017rwz} with the difference about the light-quark and strange quark.

\begin{table}[t]
\begin{center}
\renewcommand{\arraystretch}{1.5}
\footnotesize
\caption{The first tenth-order $\xi$-moments of $D_s^*$-meson longitudinal twist-2 distribution amplitude $\phi_{2;D_s^*}^\|(x,\mu )$ at the scale $\mu_0=1.0~\mathrm{GeV}$ and $\mu_k=3.0~\mathrm{GeV}$.}
\label{table:xi2}
\begin{tabular}{l@{\hspace{45pt}}l@{\hspace{45pt}}l}
\hline
 & $\mu_0=1.0\mathrm{GeV}$ & $\mu_k=3.0~\mathrm{GeV}$\\ \hline
$\langle\xi^{\|;1}_{2;D_s^*}\rangle|_\mu$ &$-0.328_{-0.051}^{+0.041}$&$-0.258_{-0.039}^{+0.033}$\\
$\langle\xi^{\|;2}_{2;D_s^*}\rangle|_\mu$ &$+0.260_{-0.039}^{+0.045}$&$+0.179_{-0.026}^{+0.031}$\\
$\langle\xi^{\|;3}_{2;D_s^*}\rangle|_\mu$ &$-0.130_{-0.020}^{+0.016}$&$-0.104_{-0.016}^{+0.013}$\\
$\langle\xi^{\|;4}_{2;D_s^*}\rangle|_\mu$ &$+0.111_{-0.016}^{+0.018}$&$+0.083_{-0.012}^{+0.014}$\\
$\langle\xi^{\|;5}_{2;D_s^*}\rangle|_\mu$ &$-0.071_{-0.011}^{+0.009}$&$-0.057_{-0.009}^{+0.007}$\\
$\langle\xi^{\|;6}_{2;D_s^*}\rangle|_\mu$ &$+0.056_{-0.008}^{+0.009}$&$+0.045_{-0.006}^{+0.008}$\\
$\langle\xi^{\|;7}_{2;D_s^*}\rangle|_\mu$ &$-0.044_{-0.007}^{+0.006}$&$-0.035_{-0.006}^{+0.005}$\\
$\langle\xi^{\|;8}_{2;D_s^*}\rangle|_\mu$ &$+0.039_{-0.006}^{+0.007}$&$+0.031_{-0.004}^{+0.005}$\\
$\langle\xi^{\|;9}_{2;D_s^*}\rangle|_\mu$ &$-0.027_{-0.004}^{+0.003}$&$-0.023_{-0.004}^{+0.003}$\\
$\langle\xi^{\|;10}_{2;D_s^*}\rangle|_\mu$ &$+0.028_{-0.004}^{+0.005}$&$+0.022_{-0.003}^{+0.004}$\\
\hline
\end{tabular}
\end{center}
\end{table}

\section{Numerical Analysis}\label{Sec:III}
For the subsequent numerical calculations, we adopt the following parameters: the mass of $m_{D_s^*}$-meson is $m_{D_s^*}=2.1122\pm 0.0004~\mathrm{GeV}$, $c$-quark current quark mass is $\bar{m}_c(\bar{m}_c) =1.28\pm 0.03~\mathrm{GeV}$, the mass of $s$-quark is $\bar{m}_s(2~\mathrm{GeV}) =0.0934_{-0.0034}^{+0.0086}~\mathrm{GeV}$, and the decay constant of $m_{D_s^*}$-meson is $f_{D_s^*}=f_{D_s^*}^\|=0.279\pm 0.019~\mathrm{GeV}$~\cite{Pullin:2021ebn}. Sometimes, when we need to evolve parameters to any other scale, we have to employ renormalization group equations (RGEs), and among them, certain inputs correspond to specific RGEs~\cite{Yang:1993bp, Hwang:1994vp}.
\begin{align}
\bar{m}_c(\mu) &=\bar{m}_c\left( \bar{m}_c \right) \left[ \frac{\alpha_s(\mu)}{\alpha_s\left( \bar{m}_c \right)} \right] ^{\frac{12}{25}},
\nonumber\\
\bar{m}_s(\mu) &=\bar{m}_s\left( 2\mathrm{GeV} \right) \left[ \frac{\alpha_s(\mu)}{\alpha_s\left( 2\mathrm{GeV} \right)} \right] ^{\frac{12}{27}}.
\end{align}

Following by the usual convention, we set the scale near the Borel parameter, {\it i.e.}, $\mu=M$. For the continuous threshold $s_0^{D_s^*}$, we require a reasonable Borel window in Eq.~(\ref{eq:Xi}) to normalize $\langle \xi_{2;D_s^*}^{\|;0}\rangle|_{\mu_0}=1$, such that the value of the continuum threshold is $s_{0}^{D_s^*}\simeq 6.8~\mathrm{GeV^2}$.

\begin{figure}[t]
\begin{center}
\includegraphics[width=0.38\textwidth]{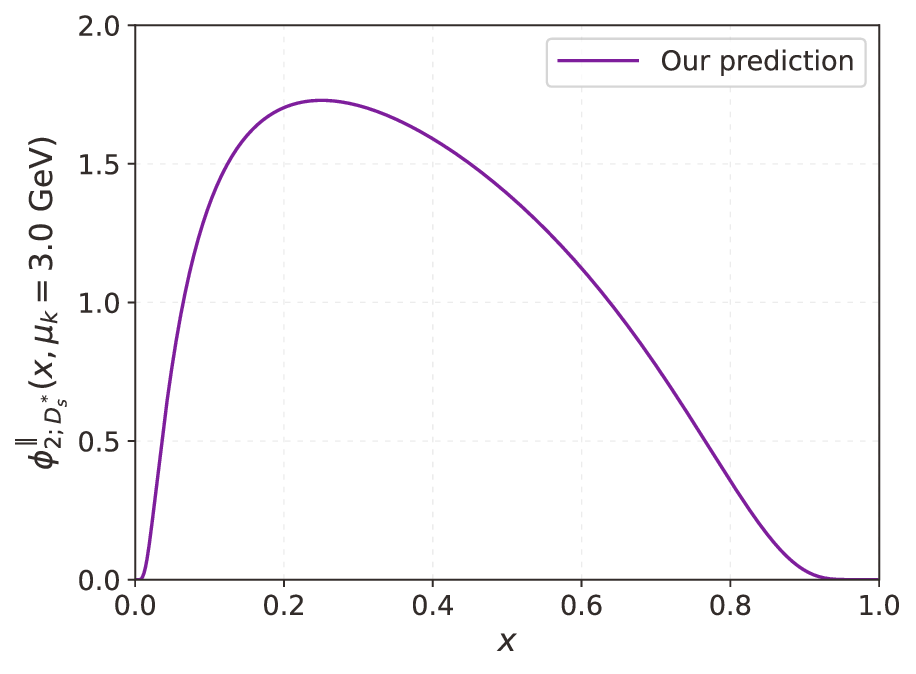}
\end{center}
\caption{The $D_s^*$-meson longitudinal leading-twist DA curves of our prediction at typical process scale $\mu_k = 3.0~{\rm GeV}$.}
\label{Fig:DA}
\end{figure}

After obtaining these input parameters, we can calculate the value of the $\xi$-moment by substituting them into the sum rule~\eqref{eq:Xi}. In this article, we calculated the values of the first ten orders of the $\xi$-moments. In order to get a stable and appropriate Borel windows for the LCDA moments, we can use the usual criterion that the contributions from continuum states. That is we take the continuum contributions do not exceed $10\%,20\%,25\%,35\%,40\%$ for the odd-order of $n=(1,3,5,7,9)$, respectively. Then we can get the upper limits of the corresponding Borel windows. For the lower limits Borel window we consider the contribution from dimension-six condensates are lower than $5\%$. The same conditions are also used in the even-order moments. Finally, we present the $D_s^*$-meson longitudinal leading-twist LCDA moments with $n=(2,4,6,8,10)$ and $n=(1,3,5,7,9)$ versus the Borel parameter $M^2$ in Fig.~\ref{Fig:fp1}(a) and Fig.~\ref{Fig:fp1}(b), separately. In which, the shaded bands are stand for the Borel windows. Then numerical results for the first ten order moments of $D_s^*$-meson twist-2 LCDAs at initial scale $\mu_0=1.0~\mathrm{GeV}$ and typical process scale $\mu_k = 3.0~{\rm GeV}$ are given in Table~\ref{table:xi2}. 

Then, using the least squares method to fit the values of the $\xi$-moments in Table~\ref{table:xi2} with the LCHO model shown in Eq.~\eqref{eq:DA}, the behavior of the $D_s^*$-meson longitudinal twist-2 LCDA $\phi_{2;D_s^*}^\|(x,\mu)$ can be determined. Through the fitting method detailed in Refs.~\cite{Zhong:2021epq,Zhong:2022ecl}, we determined the model parameters and corresponding goodness of fit. That is, by the magnitude of the probability $P_{\chi_{\min}^{2}}=\int_{_{\chi_{\min}^{2}}}^{\infty}{f(y;n_d)}dy$ with the probability density function of $\chi^2(\theta)$, $f(y;n_d) =\frac{1}{\Gamma(n_d/2) 2^{n_d/2}}y^{n_d/2-1}e^{-y/2}$, one can intuitively judge the goodness of fit. Here, $n_d$ represents the number of degrees of freedom. The optimal values of model parameters $\alpha_{2;D_s^*}$, $B_{1}^{2;D_s^*}$, and $\beta_{2;D_s^*}$ at scale $\mu_k = 3.0~\mathrm{GeV}$ and their goodness of fit are shown as follows
\begin{align}
& A_{2;D_s^*}(\mu_k) = 9.421~\mathrm{GeV^{-1}};\nonumber\\
& \alpha_{2;D_s^*}(\mu_k) = -0.860;\nonumber\\
& B_{1}^{2;D_s^*}(\mu_k) = 0.025;\nonumber\\
& \beta_{2;D_s^*}(\mu_k) = 0.774~\mathrm{GeV};\nonumber\\
& \chi_{\min}^{2}(\mu_k) = 2.463;\nonumber\\
& P_{\chi_{\min}^{2}}(\mu_k) = 92.98\%.
\end{align}
Then, the $D_s^*$-meson longitudinal leading-twist LCDA can be determined. In order to display the behavior of $\phi_{2;D_s^*}^\|(x,\mu)$ clearly and visually, we plot its curve in Fig.~\ref{Fig:DA}.

Based on the resultant $D_s^*$-meson longitudinal leading-twist DA, we can then calculate the $B_s\to D_s^*$ TFFs $A_{1}(q^2)$, $A_{2}(q^2)$, $V(q^2)$. The basic input parameters are mentioned at the beginning of this section. Therefore, the main task is to determine the continuum threshold parameter $s_0$ and the Borel window $M^2$. Based on the basic idea and process of LCSR, we have adopted the following three guidelines.

\begin{itemize}
\item The continuum contributions are less than $30\%$ of the total results;
\item The contributions from higher-twist DAs are less than $5\%$;
\item Within the Borel window, the changes of TFFs does not exceed $10\%$ ;
\item The continuum threshold $s_0^{D_s^*}$ should be closer to the squared mass of the first excited state $D_s^*$-meson.
\end{itemize}

Based on the above criteria, we determined the values of the continuum threshold $s_0^{D_s^*}$ and the Borel window $M^2$. According to the standard LCSR process, we calculated the final results of the TFFs and presented them in Table~\ref{table:fq}. For comparison, we also give the results predicted by other theoretical groups CCQM~\cite{Soni:2021fky}, QCDSR~\cite{Blasi:1993fi}, RQM~\cite{Faustov:2012mt}, PQCD~\cite{Hu:2019bdf}, LFQM~\cite{Li:2010bb}. By comparison, we can see that our predicted results are in good agreement with those predicted by these theoretical groups within the error range.

\begin{table}[t]
\begin{center}
\renewcommand{\arraystretch}{1.5}
\footnotesize
\caption{The $B_s\to D_s^*$ TFFs at large recoil region, {\it i.e.} $A_{1}(0)$, $A_{2}(0)$ and $V(0)$ with uncertainties. Meanwhile, the comparison from theoretical groups are also given.}
\label{table:fq}
\begin{tabular}{lllll}
\hline
~~~~~~~~~~~~~~~~~~~~~~&$V(0)$~~~~~~~~~~~~~~~~~~~~~~&$A_1(0)$~~~~~~~~~~~~~~~~~~~~~~&$A_2(0)$\\
\hline
This work                  &$0.647_{-0.069}^{+0.076}$  &$0.632_{-0.135}^{+0.228}$ &$0.706_{-0.092}^{+0.109}$\\
CCQM~\cite{Soni:2021fky}   &$0.743\pm0.030$            &$0.681\pm0.065$           &$0.630\pm0.025$\\
SR~\cite{Blasi:1993fi}     &$0.63\pm0.05$              &$0.62\pm0.10$             &$0.75\pm0.07$\\
RQM~\cite{Faustov:2012mt}  &$0.95\pm0.02$              &$0.70\pm0.01$             &$0.75\pm0.02$\\
PQCD~\cite{Hu:2019bdf}     &$0.64\pm0.12$              &$0.50\pm0.09$             &$0.53\pm0.11$\\
LFQM~\cite{Li:2010bb}      &$0.74_{-0.05}^{+0.05}$     &$0.61_{-0.04}^{+0.04}$    &$0.59_{-0.04}^{+0.04}$\\
\hline
\label{table:fq2}
\end{tabular}
\end{center}
\end{table}

\begin{figure}[t]
\begin{center}
\centering
\includegraphics[width=0.38\textwidth]{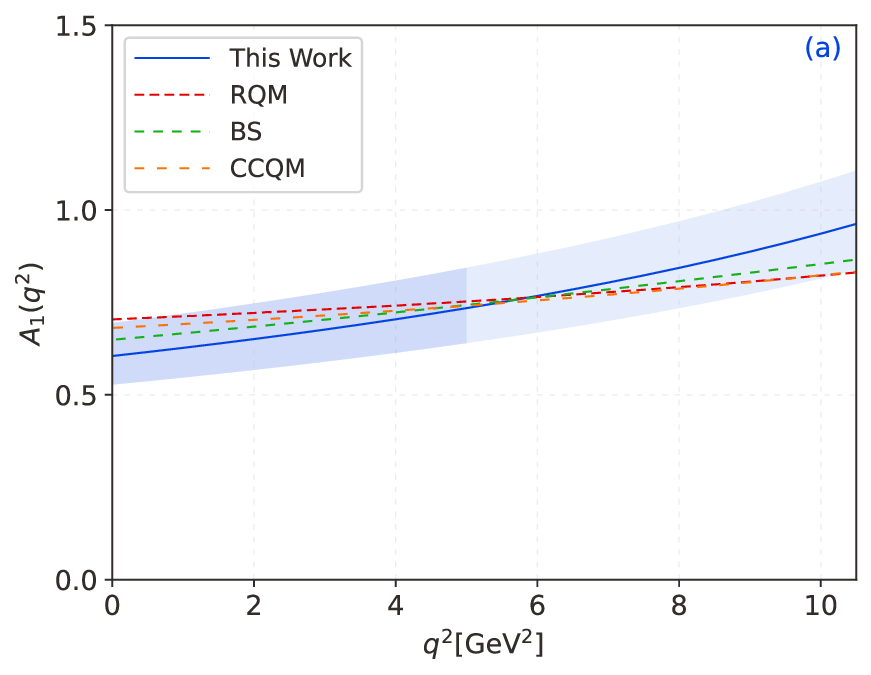}
\includegraphics[width=0.38\textwidth]{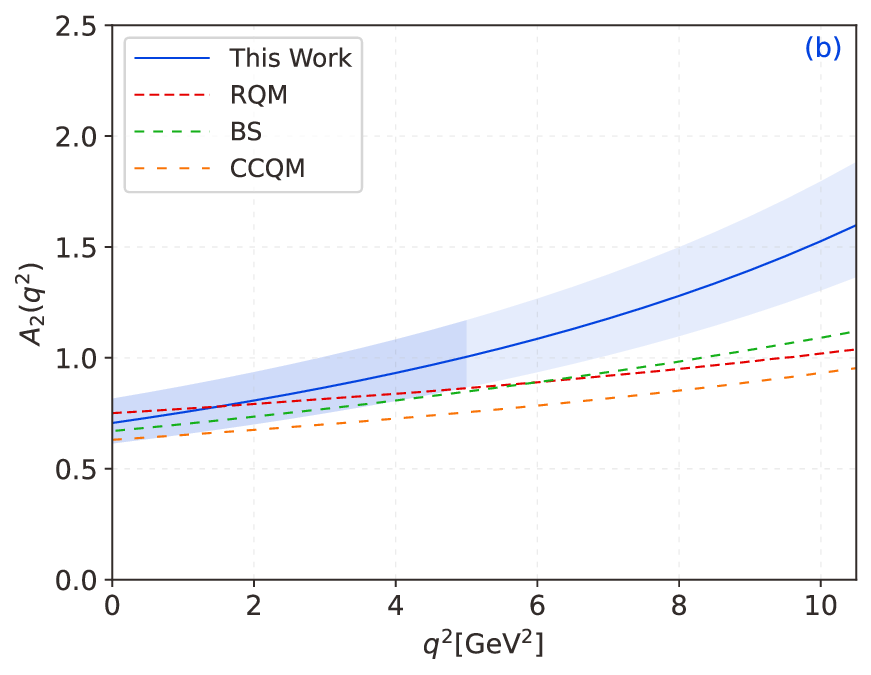}
\includegraphics[width=0.38\textwidth]{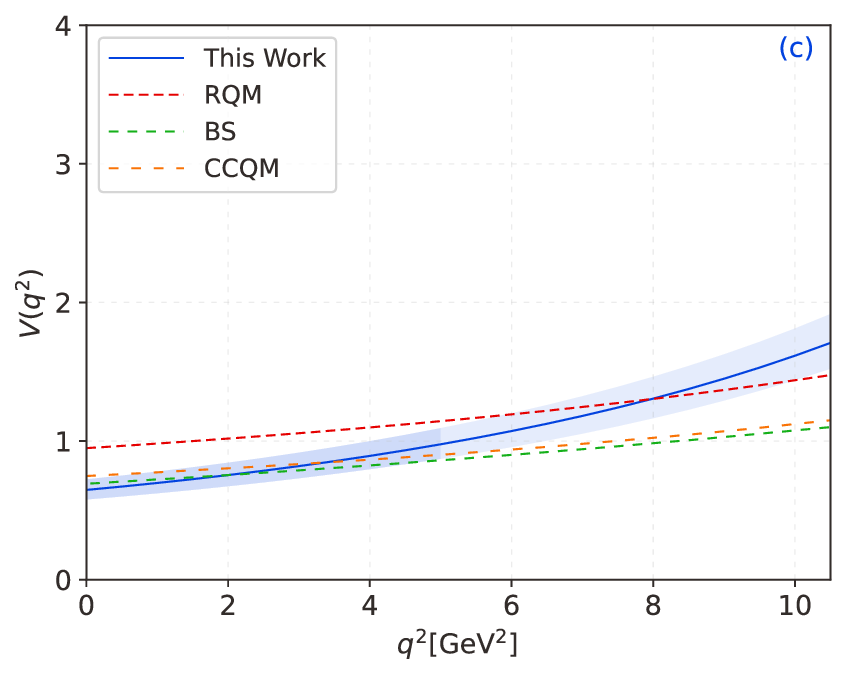}
\end{center}
\caption{The extrapolated TFFs $A_{1}(q^2)$, $A_{2}(q^2)$, $V(q^2)$ versus $q^2$. The solid blue line is the central value, and the shaded band is the corresponding uncertainty. As a comparison, the predictions
 under the RQM~\cite{Faustov:2012mt} approach, the BS~\cite{Zhou:2019stx} approach, and the CCQM~\cite{Pandya:2023ldv} approach  are also presented }
\label{Fig:fp2}
\end{figure}

In order to access  the information of $B_s\to D_s^*\ell \bar\nu_{\ell}$ TFFs in the whole kinematic region, we need to extrapolate the LCSR predictions obtained above toward large momentum transfer with a certain parametrization for the TFFs. The physically allowable ranges for the TFFs are $0 \leq q^2\leq q^2_{\rm max} = (m_{B_{s}} - m_{D_s^*})^2\sim10.6~{\rm GeV^2}$. Theoretically, the LCSR approach for $B_s\to D_s^*\ell \bar\nu_{\ell}$ TFFs is in low and intermediate $q^2$-regions, {\it i.e.} $0 \leq q^2\leq 5~{\rm GeV^2}$ of $D_s^*$-meson. One can extrapolate it to whole $q^2$-regions via a rapidly $z(q^2,t)$ converging the simplified series expansion (SSE), {\it i.e.} the TFFs are expanded as~\cite{Bharucha:2015bzk,Bharucha:2010im}:
\begin{eqnarray}
f_+(q^2) =\frac{1}{1-q^2/m_{B_{s}^2}}\sum_{k=0,1,2}{\beta_kz^k( q^2,t_0 )}
\end{eqnarray}
where $\beta_k$ are real coefficients and $z(q^2,t)$ is the function,
\begin{eqnarray}
z^k( q^2,t_0 ) =\frac{\sqrt{t_+-q^2}-\sqrt{t_+-t_0}}{\sqrt{t_+-q^2}+\sqrt{t_+-t_0}},
\end{eqnarray}
with $t_{\pm} = (m_{B_s} \pm m_{D_s^{*}})^2$ and the auxiliary parameter $t_0=t_{\pm}(1-\sqrt{1-t_-/t_+})$. The SSE method possesses superior merit, which keeps the analytic structure correct in the complex plane and ensures the appropriate scaling, $f_+(q^2)\sim 1/q^2$ at large $q^2$. And the quality of fit $\Delta$ is devoted to take stock of the resultant of extrapolation, which is defined as
\begin{eqnarray}
\Delta =\frac{\sum_t{| F_i(t) -F_{i}^{\mathrm{fit}}( t ) |}}{\sum_t{| F_i(t) |}}\times 100.
\end{eqnarray}

\begin{figure}[t]
\begin{center}
\includegraphics[width=0.38\textwidth]{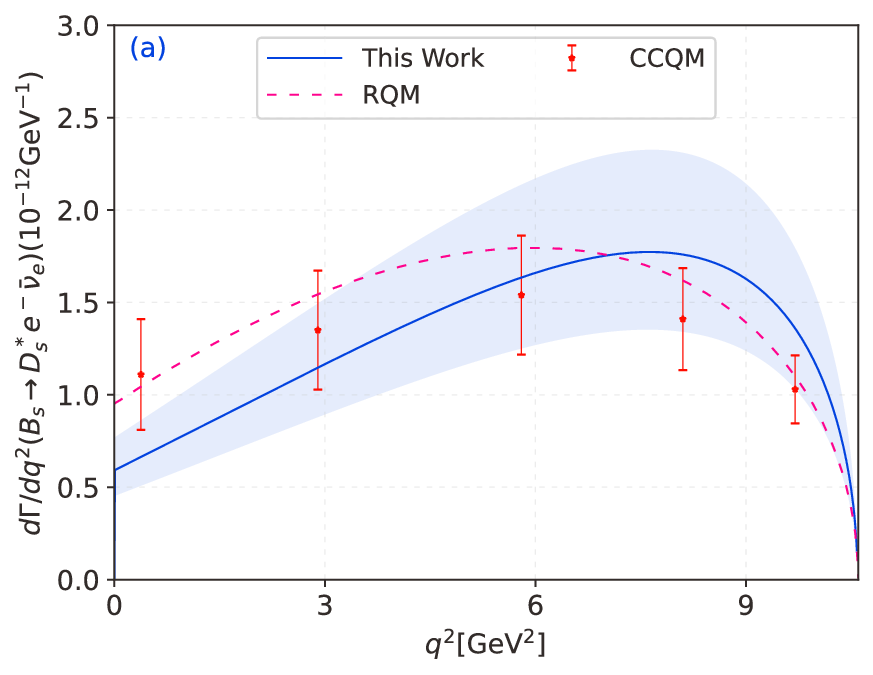}
\includegraphics[width=0.38\textwidth]{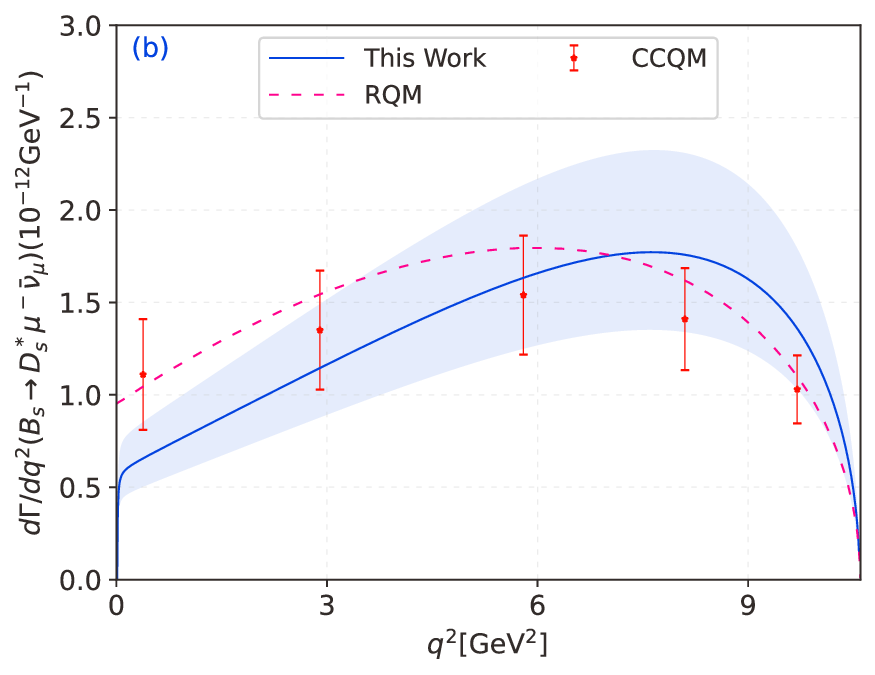}
\end{center}
\caption{Differential decay widthes for $B_s\to D_s^*\ell \bar\nu_{\ell}$ in whole $q^2$-region, where the solid line is the central value and the shaded band shows its uncertainty. As a contrast, this paper also introduces the use of different theoretical methods, such as CCQM~\cite{Soni:2021fky}, RQM~\cite{Faustov:2012mt}.}
\label{Fig:fp3}
\end{figure}

After making an extrapolation for the TFFs $A_{1}(q^2)$, $A_{2}(q^2)$, $V(q^2)$ to the physical $q^2$-region, the behaviors of the $B_s\to D_s^*\ell \bar\nu_{\ell}$ TFFs in the whole physical region can be obtained. These behaviors have been given in Fig.~\ref{Fig:fp2}. For comparison, we also present the results predicted by other theoretical groups for the TFFs, which are illustrated in detail in Fig.~\ref{Fig:fp2} below.
\begin{itemize}
\item In Fig.~\ref{Fig:fp2}, the lighter band represents the LCSR results of our prediction, while the darker band represents the SSE predictions ;
\item In the figure, the solid blue line represents this work. The red dotted line, the green dotted line, and the yellow dotted line represent the results predicted by RQM~\cite{Faustov:2012mt}, CCQM~\cite{Pandya:2023ldv}, and Bethe-Salpeter method (BS)~\cite{Zhou:2019stx} respectively;
\item In order to see the behavior of TFFs $A_{1}(q^2)$, $A_{2}(q^2)$, $V(q^2)$ more directly, as a comparison, we also show the predictions of theoretical groups such as RQM, CCQM, BS. In comparison, our results are found to be consistent within the appropriate margin of error.
\end{itemize}

The differential branching ratio is a very important physical observable measurement of the semileptonic decay process, and its study helps us to understand the internal structure of particles and their interactions with other particles. Therefore, it is very important to calculate $B_s\to D_s^*\ell \bar\nu_{\ell}$ semileptonic decay and its branching ratio. Here we use $|V_{cb}|=0.041$ to calculate~\cite{LHCb:2020cyw}, and using the branching ratio formula, we can obtain the numerical result of the branching ratio. In addition, we also present the attenuation width diagram, as shown in Fig.~\ref{Fig:fp3}. For comparison, we also show the predictions of other theoretical groups, including CCQM~\cite{Soni:2021fky} and RQM~\cite{Faustov:2012mt}. By comparison, we can see that our results agree well with both experimental and theoretical predictions within a suitable error range.

\begin{table}[t]
\begin{center}
\renewcommand{\arraystretch}{1.5}
\footnotesize
\caption{The branching fractions (in unit: $\%$) for the decays $B_s^0 \to D_s^{*+}\ell \bar\nu_{\ell}$ with $\ell = (e,\mu)$ in this work. Meanwhile, the theoretical results are given as a comparison.}
\label{table:Br}
\begin{tabular}{lll}
\hline
Decay Channel & $B_s^0\to {D_{s}^{*+}}e^-\nu_{e}$ & $B_s^0\to {D_{s}^{*+}}\mu^- \nu_\mu$ \\\hline
This Work                 & $5.45_{-1.57}^{+2.15}$ & $5.43_{-1.57}^{+2.14}$ \\
PDG~\cite{LHCb:2020cyw}   &            -           & $5.2\pm0.5$            \\
CCQM~\cite{Soni:2021fky}  & $6.42\pm0.67$          & $6.39\pm0.67$          \\
RQM~\cite{Faustov:2012mt} & $5.3\pm0.5$            & -                      \\
LFQM~\cite{Li:2010bb}     & -                      & $5.2\pm0.6$            \\
PQCD~\cite{Hu:2019bdf}    & $4.42_{-1.00}^{+1.27}$ & -                      \\
\hline
\label{table:Br}
\end{tabular}
\end{center}
\end{table}

In Table~\ref{table:Br}, we comprehensively present the numerical results for the branching ratios of various decay channels associated with the semileptonic decay of $B_s\to D_s^*\ell \bar\nu_{\ell}$. Alongside our findings, we also showcase the numerical predictions offered by other leading theoretical groups, namely CCQM~\cite{Soni:2021fky}, RQM~\cite{Faustov:2012mt}, LFQM~\cite{Li:2010bb}, PQCD~\cite{Hu:2019bdf}, and PDG~\cite{LHCb:2020cyw}. Upon conducting a meticulous comparison among these results, we notice that there exists a certain degree of deviation between our outcomes and those reported by the other groups. This discrepancy can likely be attributed to errors in the input parameters utilized in the respective calculations. Nevertheless, it is important to emphasize that, within a reasonable and acceptable range of error, our results are deemed to be valid and reliable.

\section{Summary}\label{Sec:IV}
This article studies the semileptonic decay process of $B_s\to D_s^*\ell \bar\nu_{\ell}$ within the framework of QCDSR. First, we present the expressions for the $D_s^*$-meson longitudinal twist LCDA moments $\langle \xi_{2;D_s^*}^{\|,n} \rangle|_\mu$, and list the values of the first ten $\xi$-moments at $\mu_0 = 1.0~\mathrm{GeV}$ and $\mu_k = 3.0~\mathrm{GeV}$ in Table~\ref{table:xi2}. Subsequently, we employed the least squares method to fit the values of the tenth-order $\xi$-moments at at $\mu_k=3.0~\mathrm{GeV}$ in order to determine the corresponding model parameters. Based on the BHL method, we constructed a new model of $\phi_{2;D_s^*}^{\|}(x,\mu)$, whose behavior is constrained by normalization conditions, the probability of finding the leading Fock-state in the $D_s^*$-meson Fock-state expansion, and known Gegenbauer moments. Research shows that it is necessary to find a more accurate form of meson DA. In the study of various processes, the error caused by using meson DA as an input parameter should be considered. In order to present the behavior of the LCDA $\phi_{2;D_s^*}^{\|}(x,\mu)$ in a clear and intuitive manner, we have depicted its curve in Fig.~\ref{Fig:DA}.

Then, we calculated $B_s\to D_s^*\ell \bar\nu_{\ell}$ TFFs $A_1(q^2), A_2(q^2), V(q^2)$ by using the LCSR method. We present the numerical results of the TFFs in the maximum buffer region in Table~\ref{table:fq}. After extrapolating the LCSR results of $B_s\to D_s^*$ TFFs to the entire $q^2$-region, the differential decay width and branching ratio of the semileptonic decay $B_s\to D_s^*\ell \bar\nu_{\ell}$ are obtained, as shown in Table~\ref{table:Br} and Fig.~\ref{Fig:fp3}. It was compared with other predictions and found to be consistent with experimental data and other methods within the error range.

\section{Acknowledgments}

Tao Zhong and Hai-Bing Fu would like to thank the Institute of High Energy Physics of Chinese Academy of Sciences for their warm and kind hospitality. This work was supported in part by the National Natural Science Foundation of China under Grant No.12265009, No.12265010, the Project of Guizhou Provincial Department of Science and Technology under Grant No.ZK[2025]MS219, No.ZK[2023]024.

\end{document}